# Impact of eddy currents on the dispersion relation of surface spin waves in thin conducting magnetic films

I S Maksymov and M Kostylev

School of Physics, University of Western Australia, 35 Stirling Highway, Crawley WA 6009, Australia
E-mail: ivan.maksymov@uwa.edu.au

**Short title**: Impact of eddy currents on the dispersion relation of surface spin waves

**Abstract.** We propose a rigorous solution to a long-standing problem of the impact of eddy currents on the dispersion relation of surface spin waves propagating in thin conducting magnetic films. Our results confirm the prediction of the Almeida-Mill's exchange-free theory that the inclusion of the eddy current contribution results in a deviation of the dispersion curve for the fundamental mode from the Damon-Eshbach law and a substantial linewidth broadening in a large wave vector range. We show that the decrease in the spin wave frequency is due to an increase in the in-plane component of the dynamic magnetic field within the conducting film. The decrease in the frequency is accompanied by a drastic change in the asymmetry of the modal profiles for the waves. This effect is not observable in magneto-insulating films and therefore it is unambiguously attributed to eddy currents that appear in conducting films only. We also show that the wave vector range in which eddy currents affect the dispersion curve is strongly correlated with the value of the film conductivity. This result holds for conducting films with the thickness 10-100 nm, which are considered promising for future magnonic and spintronic applications.

**PACS codes:** 75.70.Ak, 76.50.+g, 75.75.-c, 75.30.Ds, 75.40.Gb, 75.60.Ej

## 1. Introduction

Conducting magnetic films and nano-structures are the cornerstone of modern magnonic and spintronic devices. These include, but not limited to, high-density random access memory [1], spin-torque nano-oscillators [2, 3], reprogrammable magnetic logic [4], microwave devices [5-7], gas sensors [8], and current-induced spin wave Doppler shift devices [9] that allow measuring the polarization of spin waves in conducting magnetic films.

In contrast to magneto-insulating yttrium iron garnet (YIG) films (see, e.g., [10]), the fabrication of conducting films and nano-structures made, for example, of Permalloy is compatible with the modern complimentary metal-oxide semiconductor (CMOS) technology. Permalloy exhibits the optimum combination of magnetic properties: the vanishing magnetic anisotropy and the smallest magnetic (Gilbert) damping among ferromagnetic metals. The optimal magnetic properties of Permalloy are also combined with usable optical properties such as the capability of supporting surface plasmon resonances [11]. This feature opens up a channel for the interaction between light and spin waves.

One of the fundamental differences between magneto-insulating films and conducting magnetic films is the appearance of eddy currents induced by the excitation of magnetization dynamics within conductors. The effect of eddy currents was predicted in early theoretical works on conducting magnetic structures [12-18]. However, those findings receive just a little attention because the predicted effect was not confirmed experimentally (see, e.g., works on Brillouin light scattering (BLS) measurement of the spin wave dispersion in conducting single and multilayer thin films [19-22]; for an extended discussion see Ch. 14 of [23] and references therein). The reason why the theory was not confirmed experimentally is the inability of early BLS setups to measure dispersion relations of spin waves at the wave vectors smaller than $\sim 10^4$ cm$^{-1}$. This limitation is due to a very small cross-section of the inelastic scattering of photons from magnons (the quanta of spin waves) as compared to the cross-section of the elastic photon scattering process [23, 24]. Consequently, only a fractional amount of the incident laser power can be detected as a BLS signal at the photodetector of the interferometer. Thus, a high contrast between the inelastically and elastically scattered photons is required. However, the contrast becomes even lower at small angles of incidence (i.e. small wave vectors).

A relatively recent work from the mid-1990s [17] presents a simple exchange-free theory of spin wave dispersion in conducting magnetic materials. This theory predicts a deviation of the spin wave dispersion law for conducting films from the Damon-Eshbach (DE) dispersion. This variation in the dispersion is accompanied by significant broadening of the linewidth ($\Delta H$) (see figure 4 in that paper) for wave vectors $k$ in the range up to $1/\delta$, where $\delta$ is the microwave skin depth for the material. As stated in [17], this range had been inaccessible with the experimental techniques contemporary to the paper. Due to the recent advance in the conventional reciprocal space BLS caused by the current interest to microscopic magnonic crystals, it is now possible to reliably access the wave number range $<10^4$ cm$^{-1}$ including the point $k = 0$ [25]. Furthermore, the recent extension of the phase-resolved real-space BLS technique [26] to the micro-focus option of BLS setups now allows one to *simultaneously* measure the frequency and the wave vector of long-wavelength spin waves in thin metallic films with unprecedented accuracy [27].

On the other hand, a new method of ferromagnetic resonance measurements called "broadband stripline FMR" [28-34] has become popular recently. The FMR method probes an important particular point in the spin wave dispersion curve $k = 0$. It has been demonstrated that the raw stripline FMR data for metallic films are strongly affected by the conductivity effects [35-38]. These effects result in highly efficient excitation of higher-order exchange standing spin wave (SSW) modes in FMR experiments because of the perfect shielding of the microwave magnetic field by the metallic films with sub-skin-depth thicknesses. The enhancement of the SSW signal is especially large for multilayer films [35, 36]. It has been theoretically predicted that the strength of the perfect shielding is strongly spin-wave wave vector dependent [36]. Furthermore, it has been demonstrated that the contribution of spin wave excitations with small non-vanishing in-plane wave vectors ($k \neq 0$) to the resonance linewidth measured with the broadband stripline FMR method may be significant [29].

All this recently acquired experimental and theoretical knowledge requires revisiting the long-standing problem of the impact of the microwave eddy currents on the spin wave dispersion in thin metallic magnetic films. Namely:
- exchange-energy contribution to the spin wave dispersion should be included in a way allowing easy extension of the theory to the technologically important case of multilayer films containing both magnetic and non-magnetic layers;
- connection of the eddy current contribution to the *dispersion* of travelling spin waves with the perfect microwave shielding effect that affects the *excitation* of the magnetization precession should be established;
- effects of the film thickness and the magnitude of the conductivity on the dispersion relation should be understood in detail.

Even for a single-layer magnetic film, the analytical theory of dipole-exchange spin waves is highly complicated [12]. Simple analytical formulas for the dispersion law cannot be derived. Consequently, numerical approaches are required to produce the final spin wave spectra. Keeping this in mind, in this work we rely on numerical calculations from the very beginning. Similarly to the treatment of magneto-insulating films in the presence of a strong dielectric permittivity [39], we will use the Green's function approach for the description of the microwave magnetic field of spin waves. Analytical expressions for these Green's functions are extremely cumbersome even for a single magnetic layer. For this reason, the case of multiple magnetic layers is virtually untreatable analytically. Therefore, we suggest a method of numerical finite-difference formulation of the Green's functions. This method potentially allows simple extension of our theory to the case of magnetic multilayers in the future.

Our approach results in a simple numerical scheme to calculate the dispersion of dipole-exchange spin waves as a solution of a matrix eigenvalue problem. Our calculations confirm the prediction of Almeida-Mill's exchange-free theory [17] that the inclusion of the eddy current contribution results in a deviation of the dispersion curve for the fundamental mode from the DE law and a substantial linewidth broadening in a large wave vector range. Calculating the profiles of the dynamic magnetic field across the film thickness allows us to explain the decrease in the spin wave frequency as due to an increase in the in-plane component of the dynamic magnetic field within the conducting film. This effect is not observable in magneto-insulating films and therefore it is unambiguously attributed to eddy currents that appear in conducting films only. We also show that the wave vector range in which eddy currents affect the dispersion curve is strongly correlated with the value of the film conductivity. This result holds for conducting films with the thickness 10-100 nm, which are considered promising for new the generation of magnonic and spintronic devices.

In order to achieve our results, in section II we use a Bubnov-Galerkin (BG) method to construct a semi-analytical theory for the spin wave excitation in conducting magnetic films. The credibility and accuracy of the BG method is convincingly confirmed by brute-force numerical simulations by means of the finite-difference time-domain (FDTD) method, which is outlined in section III. The FDTD method is virtually exact because its accuracy is limited by the available computer memory and CPU speed only. Our FDTD method [38] takes into account the realistic electric conductivity of the film, exchange constant and Gilbert damping term. Basically, we calculate spectral density plots using the Green's function obtained with the FDTD and then extract an effective dispersion relation. Almeida and Mills [17] use the same approach. However, they derive explicit expressions for the Green's functions in the exchange-free approximation. Therefore, the difference between their results and our FDTD results should be only due to the inclusion of the exchange interaction. This and other main results are presented and discussed in section IV.

**2. Semi-analytical Bubnov-Galerkin method**

The first theoretical works on the spin-wave excitation in conducting films admitted very high complexity of calculations in the presence of conductivity [12, 16]. Despite the immense progress in computational physics in last years, the simulation of spin wave excitation in conducting magnetic films is still a challenge when using modern numerical techniques (see, e.g., the discussion in [38]). In order to calculate dispersion characteristics of spin waves in a conducting magnetic film, we modify an analytic approach relying on the magnetostatic Green's function in the Fourier space used to calculate dispersion relations of magneto-insulating films [40]. The approach presented in the cited paper was developed for the analysis of magneto-insulating films and exact analytical expressions for the Green's function were derived in the magnetostatic approximation (i.e. electric permittivity $\varepsilon = 0$). Later this approach was extended to the case of the non-vanishing relative electric permittivity [41].

Basically, the analytical expressions from [41] can be used to model the presence of the conduction currents in the magnetic layer by assuming $\varepsilon = \sigma/(i\omega\varepsilon_0)$ for the metal, where $\sigma$ is the electric conductivity of the metal layer and $\omega$ is the microwave frequency. However, these expressions are too cumbersome to deliver an explicit analytical solution for the dispersion relation [39] and the resultant system of equations should be solved numerically anyway. Therefore, instead of taking this route, we calculate the Green's functions that include the contribution of the conduction currents numerically.

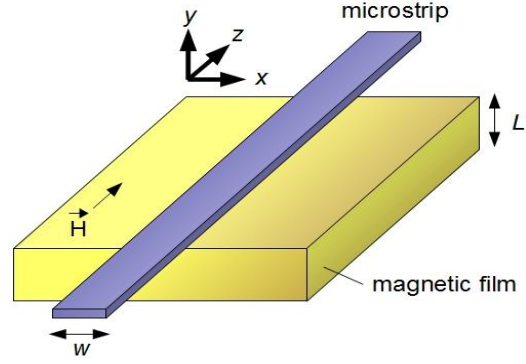

**Figure 1.** Schematic representation of a conducting magnetic film of thickness $L$ and a microstrip line of width $w$ used as a microscopic antenna exciting magnetization dynamics in the film. Note that the microstrip is present in FDTD model only, where the electric field of the microstrip serves as the driving source. In our analysis, we assume that the film is infinite in the $x$- and $z$-directions. The microstrip is infinite in the $z$-direction. The thickness of the microstrip is assumed to be negligible as compared with the thickness of the film. The conductivity of the microstrip is also neglected. The external static magnetic field is applied to the magnetic film along the $z$-axis.

*2.1. Formulation of the problem in the form of the linearized Landau-Lifshitz equation*

We consider the model in which the $y$-axis is perpendicular to the surfaces of the conducting magnetic film (figure 1). The film of thickness $L$ is continuous in the $x$- and $z$-directions. The in-plane axis $x$ coincides with the direction of the positive spin wave vector $k$. The external magnetic field $\mathbf{H}$ is applied in the positive direction of the $z$-axis. In order to describe the magnetization dynamics in the chosen model, we use the Linearized Landau-Lifshitz equation (LLLE)

$$\frac{\partial \mathbf{m}}{\partial t} = -|\gamma|(\mathbf{m} \times \mathbf{H}_0 + \mathbf{M}_0 \times \mathbf{h}_{\text{eff}}). \qquad (1)$$

In (1) the dynamic magnetization vector $\mathbf{m}$ has only two non-vanishing components ($m_x$, $m_y$) that are perpendicular to the static magnetization $|\mathbf{M}_0| = M_s$. The dynamic effective field $\mathbf{h}_{\text{eff}}$ has two components: the exchange field $\mathbf{h}_{\text{ex}}$ given by

$$\mathbf{h}_{\text{ex}} = \alpha \left( \frac{\partial^2}{\partial x^2} + \frac{\partial^2}{\partial y^2} \right) \mathbf{m} \qquad (2)$$

and the dynamic (dipole) magnetic field $\mathbf{h}$. In (2) the coefficient $\alpha$ is the exchange constant. We seek the solution of (1) in the form of a plane spin wave

$$\mathbf{m}, \mathbf{h}_{\text{eff}} = \mathbf{m}_k, \mathbf{h}_{\text{eff},k} \exp(i\omega t - kx). \qquad (3)$$

The dynamic magnetic field $\mathbf{h}$ is given by the Green's function of the electromagnetic field in the Fourier space $\mathbf{G}_k(s)$

$$\mathbf{h}_k(y) = \int_0^L \mathbf{G}_k(y - y') \, \mathbf{m}_k(y') dy' \equiv \mathbf{G}_k \otimes \mathbf{m}_k, \qquad (4)$$

where $L$ is the thickness of the magnetic film in the $y$-direction (figure 1). The LLLE includes only four components of the Green's function. For $\sigma \neq 0$ it is convenient to present the components of the function $\mathbf{G}_k(s)$ in the same form as for magneto-insulating films [40]

$$\mathbf{G}_{\mathrm{k}}(s) = \begin{pmatrix} G_{kxx} & G_{kxy} \\ G_{kyx} & G_{kyy} \end{pmatrix} = 4\pi \begin{pmatrix} -\delta(s) + G_p(k,s) & iG_q(k,s) \\ iG_q(k,s) & -G_p(k,s) \end{pmatrix}, \quad (5)$$

where $\delta(s)$ is the Dirac delta function.

In the early work where the electric permittivity was included [39] it was noted that because $G_p$ and $G_q$ are now functions of the angular frequency $\omega$ the problem cannot be formulated as an eigenvalue problem with $\omega$ playing the role of the eigenvalue. The numerical methods to solve eigenvalue problems are well established and the respective standard software is readily available [42]. Therefore, this property of $\mathbf{G}_\mathrm{k}$ for $\sigma \neq 0$ is a big disadvantage with respect to the $\sigma = 0$ case [43, 44]. Consequently, the authors of [39] had to numerically search for zeros of a complicated function of frequency and wave vector instead of straightforward solution of an eigenvalue problem.

In order to circumvent this problem, we reformulate the numerical problem such that another parameter of the LLLE plays the role of an eigenvalue. A natural choice for this parameter is the applied field, since it enters the LLLE in a linear way and it is not contained in the Green's function for the obvious reasons. We obtain

$$\omega_H \mathbf{m}_\mathrm{k} = \begin{pmatrix} G_p \omega_\mathrm{M} - \delta \omega_\mathrm{M} - \alpha k_\mathrm{n}^2 \omega_\mathrm{M} & i\omega + iG_q \omega_\mathrm{M} \\ -i\omega + iG_q \omega_\mathrm{M} & -\alpha k_\mathrm{n}^2 \omega_\mathrm{M} - G_p \omega_\mathrm{M} \end{pmatrix} \otimes \mathbf{m}_\mathrm{k}. \quad (6)$$

where $\delta$ is the Dirac delta function as above, $\omega_\mathrm{H} = \gamma H$, $\omega_\mathrm{M} = \gamma 4\pi M_\mathrm{s}$ and $k_\mathrm{n}^2 = k^2 + (n\pi/L)^2$ is the transverse wavevector with an integer $n$ being the mode number. One sees that $\omega_\mathrm{H}$ is the eigenvalue of the operator in the brackets on the right-hand side of (6). Accordingly, the eigenfunctions of this operator represent the modal profiles for the respective spin wave modes [43].

Once the numerical solution for the Green's function has been obtained, the eigenvalue problem (6) with the substituted Green's function can be solved directly in the $y$-space (i.e. in the real space) using a finite-difference approximation. However, our calculations show that the convergence of this direct-space solution of the resultant eigenvalue problem is slow [44]. A significant improvement can be achieved by using a BG method [40]. The method consists in expanding the $y$-dependence of $\mathbf{m}_\mathrm{k}$ into series using an orthonormal system of functions satisfying the exchange boundary conditions at both film surfaces. For the sake of simplicity, we assume the unpinned surface spins boundary conditions $\partial \mathbf{m}_\mathrm{k}/\partial y = 0$ at both surfaces of the magnetic film [45]. The series that satisfies these conditions is

$$\mathbf{m}_\mathrm{k}(y) = \mathbf{m}_{\mathrm{k}0} + \sqrt{2} \sum_{n=1}^{\infty} \mathbf{m}_{\mathrm{k}0} \cos\left(\frac{n\pi y}{L}\right). \quad (7)$$

By projecting (6) on the orthogonal basis of these cosine functions, we obtain an infinite system of homogeneous algebraic equations. However, in the following numerical analysis we will retain just a finite number of these equations that will be chosen to meet convergence criteria [43].

*2.2 Numerical solution for the Green's function of the Maxwell equations*

The $G_p$ and $G_q$ components of the electromagnetic field Green's function are calculated numerically by solving the Maxwell's equations using a finite-difference method with electromagnetic boundary conditions. In [35], a similar problem was solved for a multilayer film for $k = 0$. Here we demonstrate the solution for a more complicated case $k \neq 0$ for a single magnetic layer only. However, using the approach from [35] for the numerical treatment of the electromagnetic boundary conditions at the interfaces of the layers, this numerical method can be easily extended to the multilayer case. Note that both magnetic and non-magnetic layers can be treated in this way. For the non-magnetic layers in the multilayer structure one will just need to assume a vanishingly small value for the saturation magnetization, e.g. 1 Oe or so.

We seek solution in the form of (4) for the system of equations

$$\nabla \times \mathbf{h} = \sigma \mathbf{e}, \tag{8}$$

$$\nabla \times \mathbf{e} = i\omega(\mathbf{h} + \mathbf{m}), \tag{9}$$

$$\nabla \cdot \mathbf{h} = -\nabla \cdot \mathbf{m}, \tag{10}$$

where $\mathbf{m} = (m_x, m_y)$ and $\mathbf{h} = (h_x, h_y)$ as before. We make the substitution $\partial/\partial x \rightarrow -ik$, where $i$ is the imaginary unit. We obtain the following system of equations in Cartesian coordinate system

$$ikh_y + \frac{\partial h_x}{\partial y} = -\sigma e_z, \tag{11}$$

$$\frac{\partial e_z}{\partial y} = i\omega(h_x + m_x), \tag{12}$$

$$-ke_z = \omega(h_y + m_y), \tag{13}$$

$$-ikh_x + \frac{\partial h_y}{\partial y} = -\frac{\partial m_y}{\partial y} + ikm_x. \tag{14}$$

We differentiate (11) and substitute (12) into the resulting differentiated equation. Thus we obtain two cornerstone equations that entangle the components of the dynamic magnetization and magnetic field

$$\frac{\partial^2 h_x}{\partial y^2} + (i\sigma\omega - k^2)h_x = ik\frac{\partial m_y}{\partial y} - (i\sigma\omega - k^2)m_x, \tag{15}$$

$$h_y - \frac{ik\frac{\partial h_x}{\partial y}}{i\sigma\omega + k^2} = -\frac{i\sigma\omega}{i\sigma\omega + k^2}m_y. \tag{16}$$

Equations (15) and (16) must be solved consistently with the electromagnetic boundary conditions relating the electromagnetic fields inside and outside the film. This presents a difficulty because the fields outside the film can be found to constants only. These constants are the field amplitudes at the positive and negative infinities of the model in figure 1. In order to circumvent this problem, we require the magnetic field to vanish on both infinities. Also, the magnetization outside the film is zero and the medium that surrounds the film at both infinities has zero conductivity. Therefore, using (15) and (16) we find the following relations for the space *outside* the film

$$h_x = \exp(-|k|y)C, \tag{17}$$

$$h_y = -i\exp(-|k|y)h_x\frac{k}{|k|}, \tag{18}$$

where $C$ is an arbitrary constant. From the condition of continuity of $h_x$ and $e_z$ at the interfaces we obtain that $h_x^{(in)} = h_x^{(out)}$ and $e_z^{(in)} = e_z^{(out)}$, where the subscript 'in' ('out') stand for the fields inside (outside) the film. These conditions allow finding the formulas for the boundary magnetic field components *inside* the film

$$h_y = -ih_x\frac{k}{|k|}, \tag{19}$$

$$\frac{\partial h_x}{\partial y} = \frac{-i\sigma\omega}{|k|}h_x - ikh_y. \qquad (20)$$

The fact that we know the relationship between the field components at the surfaces from inside the film allows reducing the solution of the Maxwell's equations to the region inside the film only without taking care about the field values in the outside regions. The same approach was used in the work [36, 46], where good agreement between theory and experiment was found for $k = 0$.

The discrete model can be formulated in the form $\widehat{H}_{ij}^{\alpha\alpha'} h_j^{\alpha'} = \widehat{M}_{ij}^{\alpha\alpha'} m_j^{\alpha'}$, where $h_j^{\alpha}$ is the value of the dynamic magnetic field at the position $y_i$ on the mesh, $m_j^{\alpha}$ is the corresponding component of the dynamic magnetization at $y_i$, and $\widehat{H}_{ij}^{\alpha\alpha'}$ and $\widehat{M}_{ij}^{\alpha\alpha'}$ are the coefficients that arise from the discretization of (15) and (16) taking into account (19) and (20). The repeated indexes assume summation over them. As a result, the Green's function is obtained in the form $\widehat{H}^{-1}\widehat{M}$, where $\widehat{H}$ and $\widehat{M}$ are the matrices with components of $\widehat{H}_{ij}^{\alpha\alpha'}$ and $\widehat{M}_{ij}^{\alpha\alpha'}$ respectively. The discretization of $\widehat{H}_{ij}^{\alpha\alpha'}$ and $\widehat{M}_{ij}^{\alpha\alpha'}$ and the insertion of the boundary condition equations can be done similarly to the previous work [35], where this procedure is presented in much detail.

The Green's function takes the form of a matrix with the size $2n \times 2n$, where $n$ is the number of points on the one-dimensional mesh of discretization of **h** and **m** along the $y$-axis. (Recall that only the area inside the film is considered.) The numerical values of the elements of the matrix depend only on the parameter on the LHS of (15) and (16), the film thickness and the chosen mesh. Thus, this matrix is completely independent of the LLLE. Once the matrix has been obtained and projected on the basis of the cosine functions (7), it can be used in the calculations of the eigenvalues of the LLLE multiple times, e.g. for different values of $\gamma$, $4\pi M_s$, etc.

The numerical code is implemented as a MathCAD worksheet. A typical calculation of a dispersion relation, linewidth and field profiles takes about sixty seconds on a laptop computer with a single CPU and 2 GB RAM.

### 3. Finite-difference time-domain method

In order to double-check the results obtained with the semi-analytical BG method, we use a brute-force finite-difference time-domain (FDTD) method that solves the Maxwell's equations with the accuracy limited by the available CPU speed only. Previously, we developed an FDTD method that solves the Maxwell's equations consistently with the LLLE taking into account the Gilbert damping term [38]. We investigated a broadband FMR response of conducting magnetic films excited by an infinitely wide microstrip. Recall that an infinitely wide microstrip line excites spin waves with the wavevector $k = 0$ only.

Non-zero wave vectors can be excited using a finite width microstrip. Consequently, we modify the numerical model in order to take into account the value of the wave vector. The inclusion of the wave vector reduces the dimensionality of the numerical model. For this reason, the resulting algorithm is often referred to as the compact FDTD [47]. Compact FDTD methods are well-known in the theory of microwave and optical waveguides. However, in our case the numerical scheme differs from that used in other works due to the presence of the wave vector in both Maxwell's equations and LLLE. Therefore, below we derive the basic FDTD equations from the beginning in order to demonstrate the principal difference between the algorithms with and without the wave vector. We focus on only those equations that need special attention and do not present well-known results that can readily be found in the literature (see, e.g., [48-50]). We use the SI units because the Maxwell's equations written in these units are used in the major textbooks on the FDTD method (see, e.g., [48]). Also, SI system was used in our previous work [38].

We start with the Maxwell's equations for a general lossy, dispersive medium

$$\frac{\partial \mathbf{b}}{\partial t} = -\nabla \times \mathbf{e}, \qquad (21)$$

$$\varepsilon\varepsilon_0 \frac{\partial \mathbf{e}}{\partial t} = \nabla \times \mathbf{h} - \sigma\mathbf{e}, \tag{22}$$

where $\mathbf{h}$ is the dynamic magnetic field, $\mathbf{e}$ is the dynamic electric field, $\varepsilon$ is the permittivity of the medium, and $\sigma$ is the electrical conductivity. The magnetic flux density $\mathbf{b}$ is related to the dynamic magnetization in the magnetic medium through the constitutive relation $\mathbf{b} = \mu_0(\mathbf{m} + \mathbf{h})$, where $\mu_0$ is the permeability of free space. We consider the magnetic film shown in figure 1. In this case, the Maxwell's equations reduce to

$$\frac{\partial b_x}{\partial t} = -\frac{\partial e_z}{\partial y}, \tag{23}$$

$$\frac{\partial b_y}{\partial t} = -ik e_z, \tag{24}$$

$$\varepsilon\varepsilon_0 \frac{\partial e_z}{\partial t} = -ik h_y - \frac{\partial h_x}{\partial y} - \sigma e_z, \tag{25}$$

The substitution $\partial/\partial x \to -ik$ also affects the LLLE because it includes the exchange field $\mathbf{h}_{\text{ex}}$. The LLLE reads

$$\frac{\partial \mathbf{m}}{\partial t} = -|\gamma|(\mathbf{m} \times \mathbf{H}_0 + \mathbf{M}_0 \times \mathbf{h}_{\text{eff}}) - \frac{\alpha_G}{|\mathbf{M}_0|}\frac{\partial \mathbf{m}}{\partial t} \times \mathbf{M}_0, \tag{26}$$

where $|\mathbf{M}_0| = M_s$, $\mathbf{h}_{\text{eff}} = \mathbf{h} + \mathbf{h}_{\text{ex}}$ and $\alpha_G$ is the Gilbert damping term. The exchange field is given by

$$\mathbf{h}_{\text{ex}} = \frac{2A}{\mu_0 M_s^2} \nabla^2 \mathbf{m}. \tag{27}$$

Due to the previously made substitution (27) reads as

$$\mathbf{h}_{\text{ex}} = \frac{2A}{\mu_0 M_s^2}\left(\frac{\partial^2 \mathbf{m}}{\partial y^2} - k^2 \mathbf{m}\right). \tag{28}$$

The dynamic magnetization $\mathbf{m}$ and magnetic field $\mathbf{h}$ have to fulfil boundary conditions at each interface of magnetic nano-structures. It is important to notice that within the framework of the FDTD method the electromagnetic boundary conditions for the magnetic and electric fields are automatically satisfied. This is a significant advantage over the competing finite-difference techniques and it is achievable thanks to the special field arrangement in the staggered finite-difference mesh typical of the FDTD-family methods [48]. This advantage significantly increases the importance of the FDTD method for the analysis of conducting magnetic films. For instance, if the boundary conditions have to be posed manually as required by conventional finite-difference approaches [35], the simultaneous presence of conductivity and wave vector terms does not allow for separating the components of dynamic magnetic and electric fields and magnetization. This may lead to numerical instability, loss of accuracy or very poor convergence, as was observed in some of our semi-analytical calculations when $\sigma\omega/k^2 \approx 1$. We note that the BG approach significantly improves the stability with respect to the direct solution of (1)–(4) in the real space.

Unlike the electromagnetic boundary conditions, the magnetic boundary conditions for the dynamic magnetization should be posed manually at the surfaces of the film. For the most common case of the unpinned surface spins, the values of the dynamic magnetization at the interfaces should satisfy the Rado-Weertman condition that can be written as [45]

$$\frac{\partial \mathbf{m}}{\partial \mathbf{n}} = 0, \tag{29}$$

where **n** is the vector normal to the surface of the magnetic material.

The discretization and iterative solution of (21)–(29) by means of the FDTD method are similar to the procedure described in [38]. The only observation to make here is that the Courant formula for the step size in time $\Delta t$ should be a function of the wave vector of spin waves. As was shown in the previous works on the compact FDTD (see, e.g., [47]), the Courant formula should read

$$v_g \Delta t \leq \frac{1}{\sqrt{\left(\frac{k}{2}\right)^2 + \left(\frac{1}{\Delta y}\right)^2}}, \tag{30}$$

where $v_g$ is the highest speed of propagation of electromagnetic waves in any of the materials present in the model and $\Delta y$ is the discretization step along the *y*-direction. The choice of this parameter will be discussed below.

The FDTD algorithm is implemented as a FORTRAN90 source code. In contrast to the BG method, the FDTD simulations require a large computation time. For example, the calculation of one point of a dispersion relation takes more than 24 hours on a laptop computer with a single CPU and 2 GB RAM when Intel FORTRAN90 compiler is used. As noted above, the large computation time is due to a poor convergence of the simulations in the real space. However, these calculations require a very small amount of memory (~2 MB) as compared with the BG method because the dimensionality of the FDTD model is reduced due to the presence of the wave vector of spin waves in the algorithms. This means that the calculation can be accelerated by using a modern workstation with a multicore CPU. Another advantage is that the FDTD algorithm can be parallelized and used in a supercomputer or a computer cluster. Together with low memory requirements the parallelization will result in a dramatic reduction of the computation time.

**4. Results and discussion**

To start with, we calculate the dispersion relation of surface spin waves in a 100 nm-thick conductive Permalloy film (figure 1) with the following realistic material parameters: saturation magnetization $4\pi M_s = 10500$ G, exchange constant $A = 10^{-6}$ erg/cm ($\alpha = 2.28 \cdot 10^{-13}$ cm$^2$), and conductivity $\sigma = 4.5 \cdot 10^6$ S/m. The static tangential magnetic field is applied in the *z*-direction as shown in figure 1. The driving microwave frequency is 18 GHz. The gyromagnetic constant $\gamma = 2.92$ MHz/Oe is used in all calculations. The Gilbert damping constant of Permalloy is of 0.008 (used in the FDTD simulations only). We use both semi-analytical BG method and brute-force FDTD method presented in the previous sections. The direct comparison of the results obtained with the two methods allows us to validate the numerical algorithms and verify their accuracy. We also calculate the dispersion relation of surface spin waves in the magneto-insulating film the same numerical code in the limit $\sigma = 0$. It is worth noting that for $\sigma = 0$ and wave vectors smaller than $10^5$ cm$^{-1}$ the BG method demonstrates excellent quantitative agreement with the predictions of the exchange-free DE formula for the dispersion of surface spin wave modes [51].

The direct comparison of the dispersion relations in a conducting film and a magneto-insulating film reveals the impact of eddy currents induced in the Permalloy film. We observe that the fundamental mode dispersion curve (the upper solid line in figure 2) deviates from the dipole exchange DE mode of the magneto-insulating film (dashed line) towards the region of higher applied magnetic field (which corresponds to lower microwave frequencies when the applied field is fixed). For the considered 100 nm-thick Permalloy film this deviation is noticeable at the wave vectors of up to $2 \cdot 10^4$ cm$^{-1}$. As also shown in figure 2 by the dotted line, the deviation is also accompanied by a broadening of the linewidth $\Delta H$, which is obtained from the imaginary part of the complex $\omega_H$ in (6). Note that the linewidth broadening also tails off at around $2 \cdot 10^4$ cm$^{-1}$. It is important to mention that around this wave vector value the ratio $\sigma\omega/k^2$ becomes close to unity. Therefore, the increase in the wave vector dramatically affects (15) because this equation has the term $i\sigma\omega - k^2$ as its coefficients.

This result is in full qualitative agreement with the exchange-free theory [17]. Indeed, figure 4 in that paper demonstrates a similar deviation of the dispersion for $\sigma \neq 0$ from the DE law and a similar peak in $\Delta H$ at the respective frequency.

The dispersion curves of the higher-order exchange SSW modes are nearly flat (figure 2). They coincide with the dispersion curves of the higher-order SSW modes of the magneto-insulating film to graphical accuracy. The linewidth $\Delta H$ of the higher-order exchange SSW modes is negligibly small as compared with $\Delta H$ of the fundamental mode.

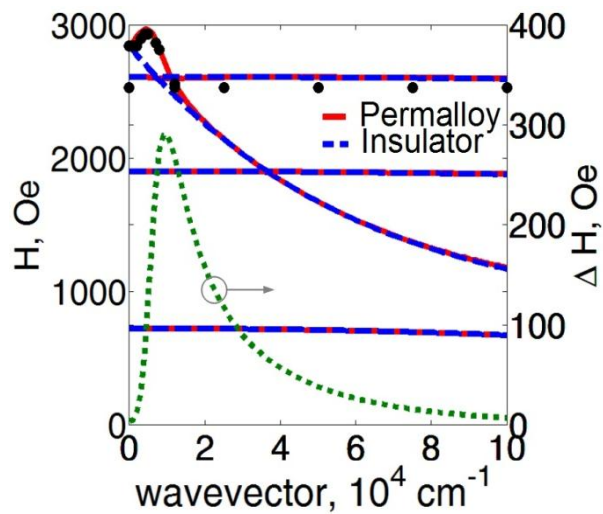

**Figure 2** The dispersion relation (solid lines) of the surface spin waves in the conducting 100 nm-thick Permalloy film ($\sigma = 4.5 \cdot 10^6$ S/m) calculated using the BG method. The driving microwave frequency is 18 GHz. The dots denote the results obtained with the FDTD method for the fundamental and the first higher-order exchange SSW modes. The dashed lines denote the dispersion relation of the 100 nm-thick magneto-insulating film ($\sigma = 0$, obtained with the BG method). The dotted line denotes the linewidth of the fundamental surface spin wave mode in the Permalloy film.

The FDTD result (thick dots in figure 2) convincingly confirms the predictions of the BG method for both fundamental and first higher-order exchange SSW modes. Our FDTD simulation did not cover the region $H < 2000$ Oe. We also note that accurate calculations of $\Delta H$ with the FDTD method are challenging due to the slow convergence of this method.

Recall that the brute-force FDTD method is virtually exact because it naturally takes into account the film conductivity, which is responsible for the induction of eddy currents, and also automatically satisfies the material boundary conditions at the metal-dielectric interfaces of the film [38]. Numerical inaccuracy of the FDTD method can only be due to (i) discretization error, (ii) numerical dispersion, and (iii) truncation of the computation domain. The truncation of the computation domain is discussed in detail in [38]. In order to minimize the influence of the other factors, we conduct numerous simulations in which we adjust the discretization for every film thickness, wave vector range and driving microwave frequency of interest. In particular, the results in figure 2 (thick dots) are obtained by retaining 20 nodes of the uniform finite-difference mesh within the 100 nm-thick Permalloy film. As can be seen from figure 2, this resolution suffices for good agreement between the results obtained with the FDTD and BG methods. Basically, accurate simulations of the dispersion curve of the first higher-order SSW require a higher mesh resolution because the spatial profile of the corresponding dynamic magnetization is more complex. However, the 20 nodes per film thickness are still good enough to identify the location of the corresponding dispersion curve with respect to that of the fundamental mode.

Here one also has to note that, as stated in [17], the correct way of constructing the spin wave dispersion in the presence of losses (in our case of eddy-current losses) is to calculate a set of spectral density functions and then to extract the dispersion relation from the positions of the peak maxima of the functions. This is exactly the way how we obtain the dispersion relation from the raw FDTD results. Thus, the excellent agreement of the BG data with FDTD data shows that in our case there is no difference between the two methods of calculation of the dispersion: through the solution of the eigenvalue problem for a homogeneous system of differential equations and through the solution of the wave excitation (spectral density) problem.

Whereas deeper understanding of the deviation of the fundamental mode dispersion curve requires a deeper theoretical analysis that will be presented below, the linewidth broadening seen in figure 2 can be explained from the point of view of the general electromagnetism. The linewidth broadening is due to energy dissipation by eddy currents induced in the conductive film by the magnetization dynamics. Eddy currents circulate inside the conducting film in such a way that a new microwave magnetic field is generated. This field contributes to the total dynamic magnetic field in the film, as will be discussed below. However, due to the internal resistance of the film eddy currents dissipate into heat, causing a removal of energy from the system that leads to the linewidth broadening.

In order to confirm the fulfilment of the laws of electromagnetism, we fix the thickness of the film to be 100 nm and vary the conductivity. As shown in figure 3, the deviation of the fundamental mode dispersion curve and the linewidth decrease rapidly as the conductivity of the film is decreased. When the conductivity of the film constitutes 1% of the conductivity of Permalloy, the deviation of the dispersion curve becomes negligible and the linewidth asymptotically goes to zero because in the BG method we do not take into account the Gilbert damping constant. The opposite behaviour is observed for the increased conductivity: the deviation of the dispersion curve becomes very large and the linewidth increases significantly. These results show that the change in the dispersion curve is conditioned by the ratio between the conductivity and wave vector. However, these results do not reveal the principle mechanism responsible for this change, whose discussion follows.

At this stage, it is important to stress the principal difference between the dispersion relations of the conducting magnetic films and metallized magneto-insulating films that have been largely investigated in the past. Dispersion properties of surface spin waves in thin films can be strongly affected by external physical conditions. Surface spin waves propagating in the opposite directions in a film are located close to the opposite surfaces of the film. Therefore, a metal screen located close to the magnetic film surface can essentially vary the dispersion. For instance, by grounding one of the film surfaces by a perfect conductor one can make the waves unidirectional in character [52]. The metallization of a YIG film does not change the dispersion relation qualitatively but leads to an increase in the bandwidth towards higher frequencies (the dispersion relations for the metalized and free standing films coincide at $k = 0$) [53]. This increase is lowered if there is an insulating gap between the film and the metal layer.

Note that at the wave vector $k = 0$ both our calculations and the DE formula produce the same result. The difference between the dispersion relations appears at the non-zero wave vectors and it reaches the maximum at $k = 9 \cdot 10^3$ cm$^{-1}$ for the 100 nm thick Permalloy film (figure 2). We plot the spatial profiles of the magnetization eigenmodes for the fundamental mode as well as the profiles of the dynamic magnetic field induced by the dynamic magnetization. All profiles are obtained by solving the system of equations (6, 15-20) numerically. Using the information about the magnetic field distribution and (8), we also calculate the electric field profiles inside the conducting film.

As shown in figure 4(a), the magnetization profiles of the conducting film (solid line) and the magneto-insulating film (dashed line) coincide to graphical accuracy at $k = 0$. It explains the coincidence of both dispersion curves at $k = 0$ in figure 2. At $k = 9 \cdot 10^3$ cm$^{-1}$ [figure 4(d)], however, the profiles of dynamic magnetization are different. In accord with the theory in [44], this result can be explained by the hybridization of the fundamental and the first higher-order SSW mode. The hybridization leads to a more pronounced asymmetry of the magnetization profile. The mode hybridization in the conducting film is weaker. Based on the result in [44], we argue that this is because at $k = 9 \cdot 10^3$ cm$^{-1}$ the fundamental mode of the conducting film is located farther from the first higher-order SSW mode than for $\sigma = 0$ (figure 2) which reduces dipole coupling between the two modes. (Below we will return to the modal profile asymmetry while discussing figure 6.)

The analysis of the profiles of the dynamic magnetic field and the corresponding electric field explains the decrease in the frequency (or the increase in the required applied field) for the spin wave in the conducting film. As shown in figure 4(b), at $k = 0$ the in-plane magnetic field $h_x$ is confined within the conducting film due to the impedance mismatch with the surrounding medium. In the insulating film $h_x = 0$. For $k = 0$, the profile of the out-of-plane magnetic field is trivial and it obeys the law of magnetostatics $h_y = -4\pi m_y$. The confinement of $h_x$ within the conducting film and the fact that this field is inductive in nature are very important for our following analysis. As shown in figure 4(c), the fields $h_x$ and $h_y$ give rise to an anti-symmetric electric field profile that, in accord with the Ohm's law, corresponds to an eddy current that flows forth near the one film surface and back near the other one. An increase in the wave vector distorts the profiles of $h_x$ [figure 4(e)] and $h_y$. Accordingly, the profile of the current flowing in the opposite directions on the opposite surfaces of the film is also distorted [figure 4(f)].

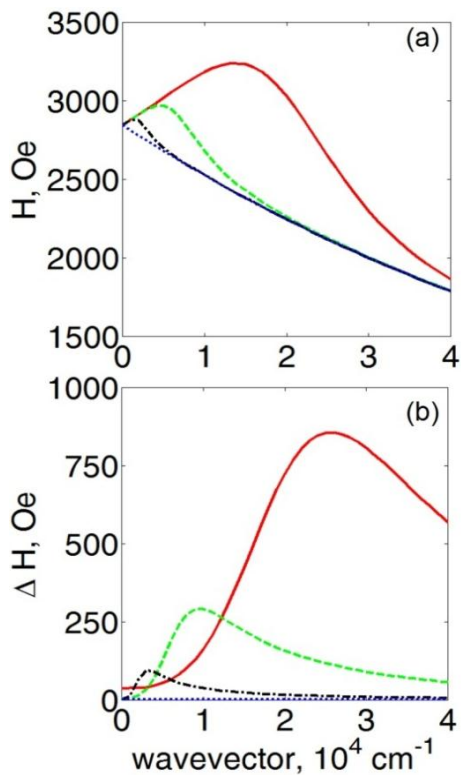

**Figure 3** The fundamental mode dispersion curve (a) and linewidth (b) of surface spin waves in 100 nm-thick conducting films of different conductivity (solid line – $4.5 \cdot 10^7$ S/m, dashed – $4.5 \cdot 10^6$ S/m, dash-dotted – $4.5 \cdot 10^5$ S/m, dotted – $4.5 \cdot 10^4$ S/m). Note that for the conductivity $4.5 \cdot 10^4$ S/m (1% of the conductivity of Permalloy) the dispersion curve coincides with that of the magneto-insulating film to graphical accuracy.

We claim that the ratio $|h_x|/|h_y|$ is a suitable parameter for quantifying the strength of the eddy-current contribution to the dispersion. The DE surface spin waves have both in-plane $h_x$ and out-of-plane $h_y$ dynamic magnetic field components. The ratio of the spatial means of the amplitudes of $h_x$ and $h_y$ is plotted in figure 5 as a function of the wave vector for the conducting film (solid line) and the magneto-insulating film (dashed line). For the magneto-insulating film the ratio $|h_x|/|h_y|$ starts from zero because at $k = 0$ for the magneto-insulating films $|h_x| = 0$. However, at $k = 0$ for the conducting films $|h_x|/|h_y| \neq 0$ (see figure 4). Most significantly, the ratio $|h_x|/|h_y|$ for the conducting films deviates from that for the insulating films precisely in the same wave vector range where the dispersion curves diverge (compare with figure 2). One sees that this range is limited from above by the special point $k = 2 \cdot 10^4$ cm$^{-1}$, where $\sigma\omega/k^2 \approx 1$.

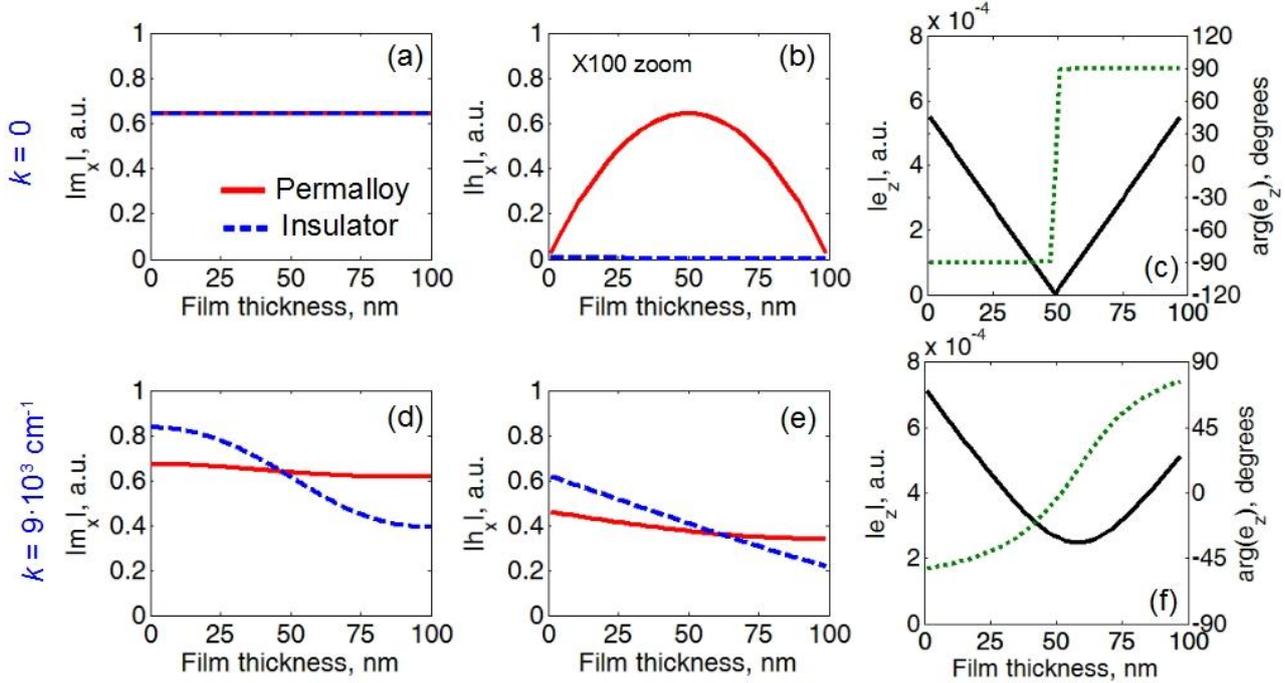

**Figure 4** Spatial profiles of the magnetization eigenmodes (a, d), dynamic magnetic field (b, e) and dynamic electric field (c, f) of the fundamental surface spin wave mode in the 100 nm thick Permalloy film (solid line) and magneto-insulating film (dashed line). The top row is for the wave vector $k = 0$ and the bottom row is for the wave vector $k = 9 \cdot 10^3$ cm$^{-1}$. Note that in the panels (c) and (f) the electric field magnitude (solid line) and phase (dashed line) are shown for the Permalloy film only. A 100-fold zoom is used in the panel (b). (Note that for $L=100$nm, $\sigma=0$, and $k = 9 \cdot 10^3$ cm$^{-1}$ the static field for the fundamental mode is smaller than for the first exchange mode (see figure 2). Therefore the asymmetry of the modal profile for $\sigma = 0$ in panel (d) is normal in contrast to Point B in figure 6. This is in contrast to the more typical example from figure 6 of a thinner film, for which the asymmetry is anomalous. This is because for thinner films the dispersion curve for the first higher-order exchange mode for $k = 9 \cdot 10^3$ cm$^{-1}$ is located lower in the field (or higher in the frequency) than the fundamental mode.)

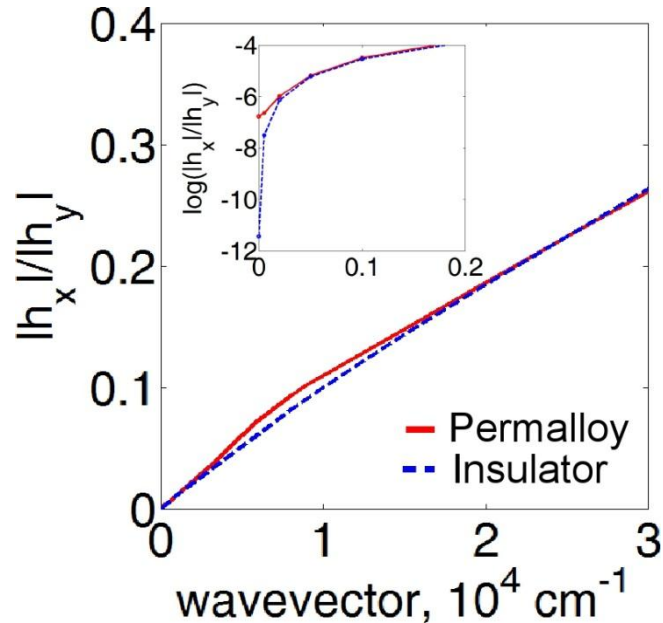

**Figure 5** The ratio of the spatial means of the amplitudes of the in-plane and out-of-plane magnetic field components as a function of the wavevector. The solid (dashed) line is for the 100 nm thick Permalloy (magneto-insulating) film. The inset shows log($|h_x|/|h_y|$) in order to demonstrate that for the Permalloy film $|h_x|/|h_y| \neq 0$ at $k = 0$. For the magneto-insulating film at $k = 0$, $|h_x|/|h_y| = 0$.

We argue that the difference between the $|h_x|/|h_y|$ for the conducting and insulating films is due to the additional dynamic in-plane field $h_x^{ec}$ induced by eddy currents in the conducting film. Thus, for the example in figure 5, the contribution of the eddy currents to the eigenfrequency reaches its maximum at around $k = 9 \cdot 10^3$ cm$^{-1}$. As follows from the comparison of figures 5 and 2, this results in the maximum divergence of the dispersion curves and the maximum for the linewidth.

The behaviour of the dynamic magnetic field profiles in the conducting films [solid lines in figures 4(b, e)] also yields additional insight into the effect of the eddy currents. Figure 4(b) suggests that $h_x$ vanishing at both film surfaces for $k = 0$ is the counterpart for the eigenwaves of the perfect shielding effect seen in the driven magnetization dynamics, when the driving microwave field is applied from one film surface only [35, 46, 54]. This shows the connection between the two effects. From figure 4(e) it can be seen that the strong non-uniformity of the field disappears as the wave vector value is increased. The field starts to be present at both film surfaces. This is in agreement with our previous prediction based on the exchange-free theory [36].

Let us now discuss in detail the effect of the conductivity on the asymmetry of the modal profiles. The left panel of figure 6 shows the dispersion relation of surface spin waves in a 50 nm-thick Permalloy film. The solid (dashed) line is the dispersion curve for the fundamental mode for $\sigma = 4.5 \cdot 10^6$ S/m ($\sigma = 0$). The dispersion curves for the first higher-order exchange SSW mode for $\sigma \neq 0$ and $\sigma = 0$ coincide to graphical accuracy. These curves are plotted in the left panel of figure 6 for the exchange constant $\alpha = 2.28 \cdot 10^{-13}$ cm$^2$ (the dash-dotted line).

Hereafter, we consider the wave vectors $k = 3 \cdot 10^4$ cm$^{-1}$ (Point A) and $k = 0.5 \cdot 10^4$ cm$^{-1}$ (Point B) denoted in the left panel of figure 6 by the vertical dashed lines. One sees that the dispersion curves for the fundamental modes of the conducting and magneto-insulating films are located higher (in terms of the $H$-field) than the curve for the exchange SSW mode at both wave vector values.

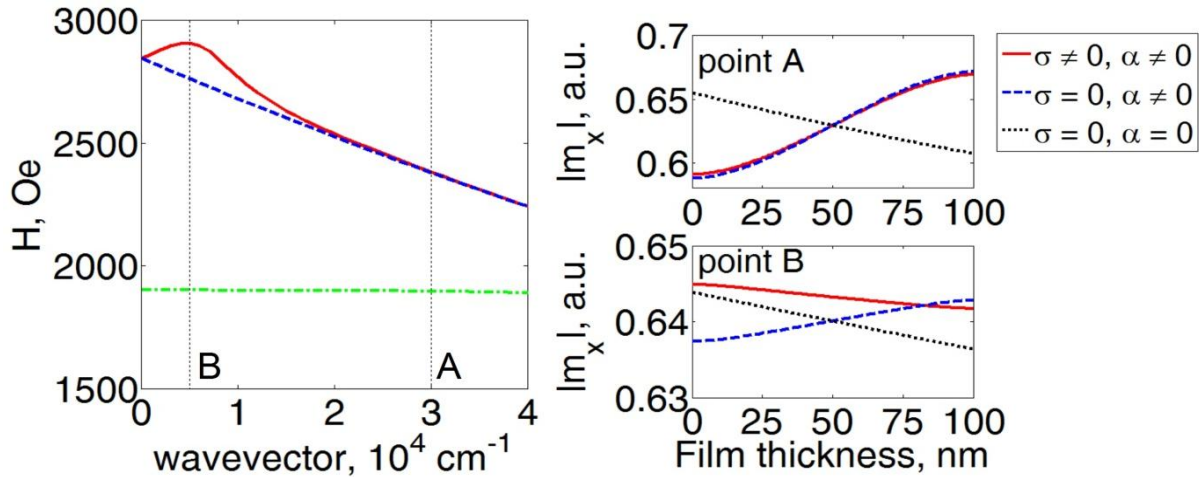

**Figure 6** Left-hand panel: the dispersion relation of surface spin waves in the 50 nm-thick Permalloy ($\sigma = 4.5 \cdot 10^6$ S/m, solid line) and magneto-insulating ($\sigma = 0$, dashed line) films. The dash-dotted line denotes the dispersion curve for the first higher-order exchange SSW for $\alpha = 2.28 \cdot 10^{-13}$ cm$^2$. The right-hand panels show the modal magnetization profiles for the points A and B. Note different *y*-axis limits used in these panels. (The profile of the exchange-free DE spin wave is given here for comparison.)

It is known that for $\sigma = 0$ the exchange interaction results in the localization of the fundamental mode at the surface opposite to the surface where the DE wave is localized (i.e. the mode localization is anomalous) [44]. This effect is seen in figure 6 (the right panel, point A) where we plot the spatial profiles of the magnetization eigenmodes at $k = 3 \cdot 10^4$ cm$^{-1}$. The respective profile for the exchange-free DE spin wave is also shown in this figure for comparison. Due to a negligible difference between the dispersion curves for $\sigma \neq 0$, and $\sigma = 0$ at the point A, a very similar profile curve is obtained for $\sigma \neq 0$. Indeed, the solid line and the dashed line are very close one to another at the point A.

However, one notices that due to the contribution of the eddy currents ($k = 0.5 \cdot 10^4$ cm$^{-1}$, point B) the profile for the fundamental mode of the dipole-exchange waves is no longer anomalous. Indeed, at the point B the profile symmetry for $\sigma = 0$ remains unchanged as compared with the point A; however, for $\sigma \neq 0$, the profile curve has the opposite symmetry, i.e. the asymmetry which is qualitatively the same as for the DE exchange-free waves. This suggests that the modal asymmetry analysis from [44] is not applicable to the case $k^2/\sigma\omega < 1$, possibly because of the different origin of the $h_x$ component of the dynamic magnetic field in the presence of conductivity. The direct implication of this result is that the potential impact of the conductivity should be kept in mind while extracting the degree of spin polarization of electrons from the results of measurements of the Doppler shift of spin waves [55, 56].

Hereafter, we demonstrate that the wave vector range in which the dispersion relation is affected by the eddy currents is independent of the thickness of the conducting film. We fix the conductivity and change the thickness of the film. We consider Permalloy films with thicknesses of several tens of nanometres. Such films are largely used in experiments and they are considered highly promising for applications in magnonics and spintronics. As shown in figure 7, the deviation of the dispersion curve form the DE law increases as increases the film thickness. The linewidth also follows this trend. Our explanation of this effect is based on (15). From (15) one sees that the effect of eddy currents should be important for the spin wave wavelengths comparable with the microwave skin depth [i.e. $k$ comparable to $(\sigma\omega)^{1/2}$]. One also sees that the case $k^2 \ll \sigma\omega$ is different from the case $k^2 \gg \sigma\omega$ and the DE dispersion case $\sigma = 0$. This difference results in the deviation of the dispersion from the DE law. This conclusion implies that the amplitude of this deviation may vary with the film thickness, as we also see in our calculation, but the wave vector range where the effect of eddy currents is important should not change.

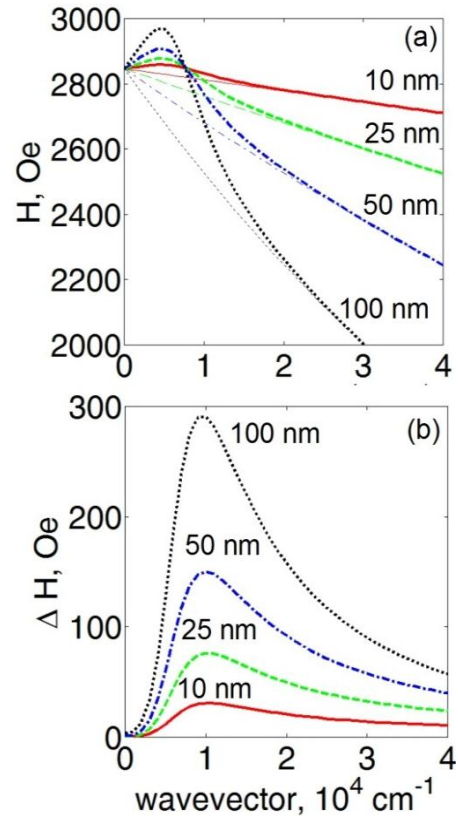

**Figure 7** The dispersion curve (a) and linewidth (b) for the fundamental mode of dipole exchange spin waves in the Permalloy films ($\sigma = 4.5 \cdot 10^6$ S/m) of different thickness (solid line – 10 nm, dashed – 25 nm, dash-dotted – 50 nm, dotted – 100 nm). In the panel (a) the thin lines denote the dispersion curves for the magneto-insulating films of the same thickness.

The dispersion relation plotted in the *H-k* domain in figure 2 is not always convenient for the interpretation of experimental results because in many cases measurements are conducted in the ω-*k* domain (see, e.g., [57-59]). By plotting a number of dispersion curves in the *H-k* domain for the 25 nm and 100 nm-thick Permalloy films and combining the results, we produce the dispersion curves in the ω-*k* domain for different values of the static magnetic field *H* (figure 8). In accord with the modification of the dispersion curve due to non-vanishing σ observed in the *H-k* domain, the resulting dispersion curves in the ω-*k* domain also exhibit a redshift. One sees that for the 100 nm-thick film the redshift reaches several GHz with respect to the dispersion curves for the magneto-insulating film. The redshift is less pronounced for the 25 nm-thick film. For both films the redshift tails off at the same value of the wavevector $k = 2 \cdot 10^4$ cm$^{-1}$.

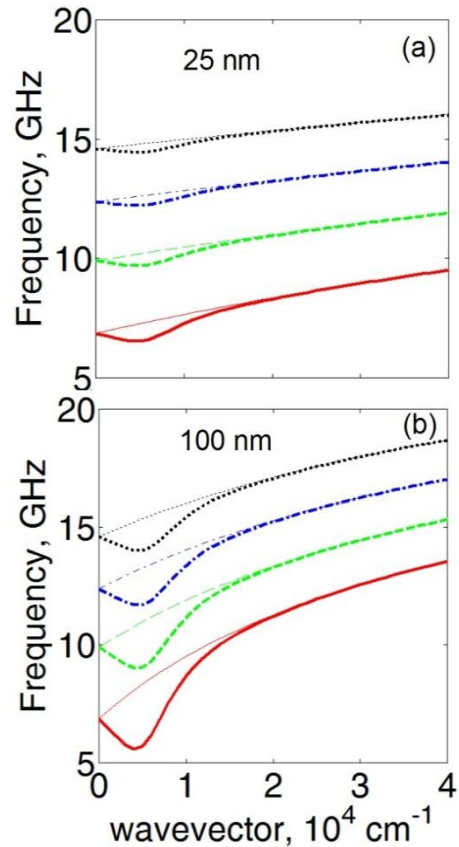

**Figure 8** The dispersion curves of the fundamental surface spin wave mode in the ω-k domain: (a) 25 nm-thick Permalloy film and (b) 100 nm-thick Permalloy film. In both panels the thick solid line denotes the dispersion relation at $H$ = 500 Oe, dashed line – 1000 Oe, dashed-dotted line – 1500 Oe, and the dotted line – 2000 Oe. The dispersion curves of the corresponding magneto-insulating films are denoted by thin lines.

The redshift shown in figure 8 may be important for correct interpretation of experimental data. For example, the micro-BLS spectroscopy allows accessing quantities such as the space and time-resolved phase profiles and the wave fronts of spin-wave packets [26]. Due to the redshift of the dispersion curve in figure 8 one can simultaneously excite two spin waves of the same frequency in the wave vector range affected by eddy currents. The interference of the two spin waves and high $\Delta H$ corresponding to this region [see figure 7(b)] can introduce uncertainty into BLS-measured space and time-resolved phase profiles of spin waves. Similarly, in the travelling spin wave inductive spectroscopy experiment [9] it will be difficult to detect these waves because of the increase spatial decay due to high $\Delta H$.

We argue that the increased losses due to eddy currents at small wave vectors may explain the gradual decrease in the amplitude the spin waves at the frequencies around the cut-off frequency (the frequency for $k$ = 0) of the Permalloy waveguide while excited with the microstrip transducer (7-8.5 GHz frequency range in figure 1(b) in [58]). We claim that the real value of the cut-off frequency should be lower as compared with the value indicated in [58]. Finally, we point out that the theoretical explanation of the linewidth broadening due to excitation of travelling spin waves in a broadband FMR experiment [29] should be revised taking into account the effect of eddy currents at small wave vectors. Indeed, the theory in the cited paper assumes that the dispersion curves are linear and with a positive slope at small wave vectors. However, this assumption disagrees with our theoretical predictions in figure 8. Furthermore, it does not include the strong increase in the magnetic losses with an increase in the wave number from zero seen in figure 7(b).

## 5. Conclusions

We constructed a semi-analytical theory of the spin wave excitation in conducting magnetic films. Our calculation approach uses the numerically calculated Green's function of the electromagnetic field. In contrast to analytical theories, the numerics-based approach makes it possible to include the exchange interaction without significant difficulties. This advantage was shown on the example of thin single-layer films. Extension of our theory to the technologically important case of multilayer nano-structures containing magnetic and non-magnetic metals is straightforward.

In good agreement with a brute-force numerical finite-difference time-domain method, we demonstrated that the appearance of eddy currents in realistic Permalloy films not only leads to the energy dissipation in the system, but also increases the in-plane component of the dynamic magnetic field. As a result, in contrast to magneto-insulating films, the in-plane magnetic field in the conducting films is always non-zero. We showed that whereas the energy dissipation by eddy currents leads to the linewidth broadening, the increase in the in-plane magnetic field due to eddy currents is the principal mechanism responsible for the deviation of the dispersion relation of surface spin waves from the DE law.

We found that the conductivity effect does not modify the dispersion of the higher-order standing spin waves: in the DE geometry they remain dispersionless for small wave vectors similar to magneto-insulating films. The effect of eddy currents is more pronounced for the thicker films as well as for the films with higher conductivity.

By analysing spatial profiles of the dynamic magnetization and the dynamic electromagnetic fields, we revealed that the in-plane microwave magnetic field is confined within the conducting film due to the impedance mismatch with the surrounding medium for $k = 0$. In order to satisfy this field distribution, eddy currents flow forth on the one film surface and back on the other one.

With the increase in the wave vector the amplitude of the microwave magnetic field at the film surfaces grows. This result is in agreement with our previous prediction based on the exchange-free theory. The modal profile asymmetry resulting in the surface character of the fundamental mode of the dipole exchange waves is drastically affected by the effect of the eddy current. Due to the eddy currents induced in the film by the precessing magnetization the asymmetry may change from anomalous to normal, i.e. the wave may become localized at the same surface as the exchange-free DE wave in the magneto-insulating materials of the same thickness. This normal asymmetry is observed for the spin wave wavelengths smaller than the microwave skin depth for the conducting material. For larger wave numbers the asymmetry is anomalous as in the absence of conductivity.

The results of our theoretical study will be of practical interest for the development of a wide range of magnonic and spintronic devices where the knowledge of the spin-wave dispersion relation is of importance. Our findings can also help to understand experimental results obtained by Brillouin light scattering spectroscopy of spin wave excited by sub-micron-wide microwave antennas in magnetic nano-structures and thin films made from conduced materials.

**Acknowledgements**
This work was supported by the Australian Research Council. ISM gratefully acknowledges a postdoctoral research fellowship from the University of Western Australia.

**References**
[1] Åkerman J 2005 *Science* **308** 508
[2] Demidov V E, Urazhdin S and Demokritov S O 2010 *Nat. Mater.* **9** 984
[3] Madami M, Bonetti S, Consolo G, Tacchi S, Carlotti G, Gubbiotti G, Mancoff F B, Yar M A and Åkerman J 2011 *Nat. Nanotech.* **6** 635
[4] Ding J, Kostylev M and Adeyeye A O *Appl. Phys. Lett.* **100** 073114
[5] Cramer N, Lucic D, Camley R E and Celinski Z 2000 *J. Appl. Phys.* **87** 6911

[6] Salahun E, Quéffélec P, Tanné G, Adenot A-L and Acher O 2002 *J. Appl. Phys.* **91** 5449.
[7] Khivintsev Y V, Reisman L, Lovejoy J, Adam R, Schneider C M, Camley R E and Celinski Z J 2010 *J. Appl. Phys.* **108** 023907
[8] Chang C S, Kostylev M and Ivanov E 2013 *Appl. Phys. Lett.* **102** 142405
[9] Vlaminck V and Bailleul M 2008 *Science* **322** 410
[10] Serga A A, Chumak A V and Hillebrands B 2010 *J. Phys. D: Appl. Phys.* **43** 264002
[11] Kostylev N, Maksymov I S, Adeyeye A O, Samarin S, Kostylev M and Williams J F 2013 *Appl. Phys. Lett.* **102** 121907
[12] Wolfram T and De Wames R E 1971 *Phys. Rev. B* **4** 3125
[13] Vittoria C, Bailey G C, Barker R C and Yelon A 1973 *Phys. Rev. B* **7** 2112
[14] Gurevich A G 1974 *Sov. Solid State Phys.* **16** 1784
[15] Maryshko M 1975 *Phys. Stat. Sol. A* **28** K159
[16] Wilts C H and Ramer O G 1976 *J. Appl. Phys.* **47** 1151
[17] Almeida N S and Mills D L 1996 *Phys. Rev. B* **53** 12232
[18] Sukstanskii A and Korenivski V 2000 *J. Magn. Magn Mater.* **218** 144
[19] Srinivasan G and Patton C E 1987 *J. Appl. Phys.* **61** 4120.
[20] Vohl M, Barnaś J and Grünberg P 1989 *Phys. Rev. B* **39** 12003
[21] Kabos P, Patton C E, Dima M O, Church D B, Stamps R L and Camley R E 1994 *J. Appl. Phys.* **75** 3553
[22] Demokritov S O and Tsymbal E 1994 *J. Phys.: Condens. Matter* **6** 7145
[23] Zhu Y 2005 *Modern Techniques for Characterizing Magnetic Materials* (Springer: Berlin)
[24] Demokritov S O, Hillebrands B and Slavin A N 2001 *Phys. Rep.* **348** 441.
[25] Tacchi S, Madami M, Gubbiotti G, Carlotti G, Goolaup S, Adeyeye A O, Singh N and Kostylev M P 2010 *Phys. Rev. B* **82** 1844081
[26] Serga A A, Schneider T, Hillebrands B, Demokritov S O and Kostylev M P 2006 *Appl. Phys. Lett.* **89** 063506
[27] Demokritov S O and Demidov V E 2008 *IEEE Trans. Magnetics* **44** 6
[28] Silva T J, Lee C S, Crawford T M and Rogers C T 1999 *J. Appl. Phys.* **85** 7849
[29] Counil G, Kim J-V, Devolder T, Chappert C, Shigeto K and Otani Y 2004 *J. Appl. Phys.* **95** 5646
[30] Schneider M L, Kos A B and Silva T J 2004 *Appl. Phys. Lett.* **85** 254
[31] Schneider M L, Gerrits T, Kos A B and Silva T J 2005 *Appl. Phys. Lett.* **87** 072509
[32] Kalarickal S S, Krivosik P, Wu M, Patton C E, Schneider M L, Kabos P, Silva T J and Nibarger J P 2006 *J. Appl. Phys.* **99** 093909
[33] Counil G, Crozat P, Devolder T, Chappert C, Zoll S and Fournel R 2006 *IEEE Trans. Magn.* **42** 3321
[34] Bilzer C, Devolder T, Crozat P, Chappert C, Cardoso S and Freitas P P 2007 *J. Appl. Phys.* **101** 074505
[35] Kostylev M 2009 *J. Appl. Phys.* **106** 043903
[36] Kennewell K J, Kostylev M, Ross N, Magaraggia R, Stamps R L, Ali M, Stashkevich A A, Greig D and Hickey B J 2010 *J. Appl. Phys.* **108** 073917
[37] Kostylev M, Stashkevich A A, Adeyeye A O, Shakespeare C, Kostylev N, Ross N, Kennewell K, Magaraggia R, Roussigné Y and Stamps R L 2010 *J. Appl. Phys.* **108** 103914
[38] Maksymov I S and Kostylev M 2013 *J. Appl. Phys.* **113** 043927
[39] Demidov V E, Kalinikos B A and Edenhofer P 2002 *J. Appl.Phys.* **91** 10007
[40] Kalinikos B A 1981 *Sov. J. Phys.* **24** 718
[41] Demidov V E and Kalinikos B A 2000 *Tech. Phys. Lett.* **26** 729
[42] Wilkinson J H and Reinsch C 1971 *Handbook for Automatic Computation: Linear Algebra* vol 2 (Springer-Verlag: Berlin)
[43] Kostylev M P, Kalinikos B A and Dötsch H 1995 *J. Magn. Magn. Mater.* **145** 93
[44] Kostylev M 2013 *J. Appl. Phys.* **113** 053907
[45] Rado C T and Weertman T R 1959 *J. Phys. Chem. Solids* **11** 315

[46] Kostylev M 2013 *J. Appl. Phys.* **113** 053908
[47] Cangellaris A C 1993 *IEEE Microwave Guided Wave Lett.* **3** 3
[48] Sullivan D M 2000 *Electromagnetic Simulation Using the FDTD Method* (IEEE Press: New York)
[49] Maksymov I S, Marsal L F and Pallarès J 2004 *Opt. Commun.* **239** 213
[50] Maksymov I S, Marsal L F, Ustyantsev M A and Pallarès J 2005 *Opt. Commun.* **248** 469
[51] Damon R W and Eshbach J R 1961 *J. Phys. Chem. Solids* **19** 308
[52] Seshadri S R 1970 *Proc. IEEE* **58** 506
[53] Adam J D 1985 *IEEE Trans. Magn.* **MAG-21** 1794
[54] Kostylev M 2012 *J. Appl. Phys.* **112** 093901
[55] Sekiguchi K, Yamada K, Seo S-M, Lee K-J, Chiba D, Kobayashi K and Ono T 2012 *Phys. Rev. Lett.* **108** 017203
[56] Haidar M, Bailleul M, Kostylev M and Lao Y *unpublished*
[57] Tsankov M A, Chen M and Patton C E 1996 *J. Appl. Phys.* **79** 1595
[58] Demidov V E, Kostylev M P, Rott K, Krzysteczko P, Reiss G and Demokritov S O 2009 *Appl. Phys. Lett.* **95** 112509
[59] Gubbiotti G, Tacchi S, Madami M, Carlotti G, Adeyeye A O and Kostylev M 2010 *J. Phys. D: Appl. Phys.* **43** 264003.